\newcounter{myctr}
\def\myitem{\refstepcounter{myctr}\bibfont\noindent\ifnum\themyctr>9\else\phantom{0}\fi\hangindent17pt\themyctr.\enskip}

\documentclass{ws-ijqi}
\usepackage{hyperref}
\usepackage[super,sort,compress]{cite}
\usepackage{graphicx}
\usepackage{caption}
\usepackage{subcaption}
\usepackage{xcolor}

\newcommand{\ii}{\mathrm{i}}
\newcommand{\ee}{\mathrm{e}}

\def\beq{\begin{equation}}
\def\eeq{\end{equation}}
\def\bea{\begin{eqnarray}}
\def\eea{\end{eqnarray}}
\def\nn{\nonumber}
\def\tr{\text{Tr}}

\begin{document}

\catchline{}{}{}{}{}

\title{ENHANCEMENT OF EFFICIENCY IN THE 
DICKE MODEL
QUANTUM HEAT ENGINE}

\author{M. Aparicio Alcalde}

\address{Instituto de Ci\^encias Exatas e Tecnol\'ogicas, Universidade Federal de Vi\c{c}osa,\\  Rio Parana\'iba - MG, CEP 38810-000, Brazil\\
martin.aparicio@gmail.com}

\author{E. Arias}

\address{Instituto Polit\'ecnico, Universidade do Estado do Rio de Janeiro,\\ Nova Friburgo - RJ, CEP 28625-570, Brazil\\
earias@iprj.uerj.br}

\author{N. F. Svaiter}
\address{Centro Brasileiro de Pesquisas F\'{\i}sicas, 22290-180 Rio de Janeiro, RJ, Brazil\\
nfuxsvai@cbpf.br}

\maketitle


\begin{abstract}
We analyze a quantum heat engine described by the full Dicke model. The system exhibit quantum phase transitions  under certain conditions.
We consider the system  performing a Stirling thermodynamic cycle. We obtain an enhancement of efficiency when during the cycle the coupling parameter cross a critical value. We analyze the effect of unbalance between rotating and counter-rotating terms in the model. 
The maximum efficiency is obtained when 
the contributions of the counter-rotating and rotating terms are equal.
The relation between ground state degeneracy, related to the quantum phase transition, and maximum efficiency is investigated.
\end{abstract}



\markboth{M. Aparicio, E. Arias and N. Svaiter}
{Quantum Heat Engine and Quantum Phase Transitions}

\section{\label{sec:level1}Introduction}	
The thermodynamics of quantum systems have been enriched and gained new perspectives by introducing the effects of quantum fluctuations, quantum correlations as well as strong system-reservoir coupling and non-Markovian dynamics.
The analysis related to this field became relevant not only for fundamental research but also for modern technological applications \cite{Millen2016, Zeeya2017, Alicki2018}.
Remarkable results using quantum resources in technology are for example the use of quantum entanglement for quantum computation and quantum information \cite{nielsen}.
In a similar way, thermal machines that use quantum systems have attracted the attention of the community. 
These kind of machines are called quantum heat engines \cite{Alicki1979, Kieu2004}.
In order to study the heat exchange and work production in such engines, the concepts of heat, work and free energy need to be reviewed in the quantum realm, therefore one have to refer to the prescriptions of quantum thermodynamics.
The principal advantage of quantum heat engines over its classical counterpart, is that higher efficiencies could be reached 
\cite{niedenzu01,lutz1,niedenzu02,squeezed01,squeezed02,squeezed03,su2018,Gardas}. Some quantum resources that allow such efficiency boost are the use of non-thermal reservoirs, as for example squeeze reservoirs and quantum-coherent baths, even quantum vacuum fluctuation could be exploited as non-standard reservoir for quantum heat engines \cite{enrique2018}. 
All this interesting phenomenology motivated the experimental realization of quantum heat engines \cite{expQHE01,expQHE02}.
%
In addition, other resource allowing quantum heat engines to increase their efficiency and power are collective effects in many-body scenario \cite{collective00, collective01,collective02,collective03,niedenzu03}. A well known collective phenomena of matter is quantum phase transition \cite{sachdev}. Therefore, recent works found connections between quantum phase transition with higher efficiencies and power increase in quantum heat engines \cite{ma,fazio,kloc}. Similarly, other study \cite{fadaie} explore the effects of topological phase transition in the work and efficiency of a quantum heat engine.

A paradigmatic quantum critical model is an ensemble of two-level system interacting with a common bosonic mode inside an optical cavity, known as Dicke model \cite{Dicke00, Dicke01,Dicke02}. The model exhibit second order quantum phase transition \cite{Aparicio2007, Aparicio2009, Aparicio2010, Aparicio2011}. 
The model reveals interesting relations between quantum phase transition, maximum quantum entanglement and quantum chaos \cite{LMGEntang01,DickeEntang01, chaos-lmg1,chaos-lmg2}. Moreover, it manifests a generalization of their criticality called of excited state quantum phase transition, which is an active area of study \cite{heiss01,LMG3,relano1,brandes}. For an experimental realization of such model see \cite{baumann,baden}.
%
%

In this work we study a quantum heat engine defined by a model that exhibit quantum phase transition. Due to its collective behavior, already proved to be important in other quantum technology set-ups \cite{fusco,QTDicke01}, we select the full Dicke model. 
We refer by full because the use of different coupling constants for the rotating and counter-rotating interaction terms. 
The quantum heat engine follow a Stirling thermodynamic cycle, defined by two isothermal processes (and changing coupling constant) and two processes at fixed coupling constant (and changing temperature). 
Higher efficiency values are obtained when during the process some parameters of the system crossover the critical values, and eventually Carnot efficiency is achieved. The quantum phase transition happens at the limit of infinite number of two-levels systems, nonetheless in our analysis also is considered finite size systems, where Carnot efficiency also is reached. Moreover, we analyze the effect of  the unbalance between the rotating and counter-rotating terms in the  model. We show that the efficiency grows   when the rotating and counter-rotating terms are equally balanced.   The main result of the work is that the presence of the counter-rotating terms  increases the efficiency. 
%
%

This paper is organized as follows. In Secction \ref{sec:level2} we introduce the full Dicke model and some properties related to the quantum phase transition. In Secction \ref{sec:level3} it is shown how to obtain the system thermodynamic quantities and the quantum heat engine is defined. In Secction \ref{sec:level4} we analyze the heat production, work and efficiency of the quantum heat engine and its relation with the quantum phase transition.
In Secction \ref{sec:level5} we set our principal conclusions and final considerations. Finally, in \ref{apdeg}  the relation between ground state double degeneracy and efficiency increase in quantum heat engines is analyze. In this work units with $\hbar=k_B=1$ are used.

\section{\label{sec:level2} The Quantum Phase Transtion in the Full Dicke Model  }
\begin{figure}[t!]
    \centering
    \begin{subfigure}{0.49\textwidth}
        \centering
        \includegraphics[width=\linewidth]{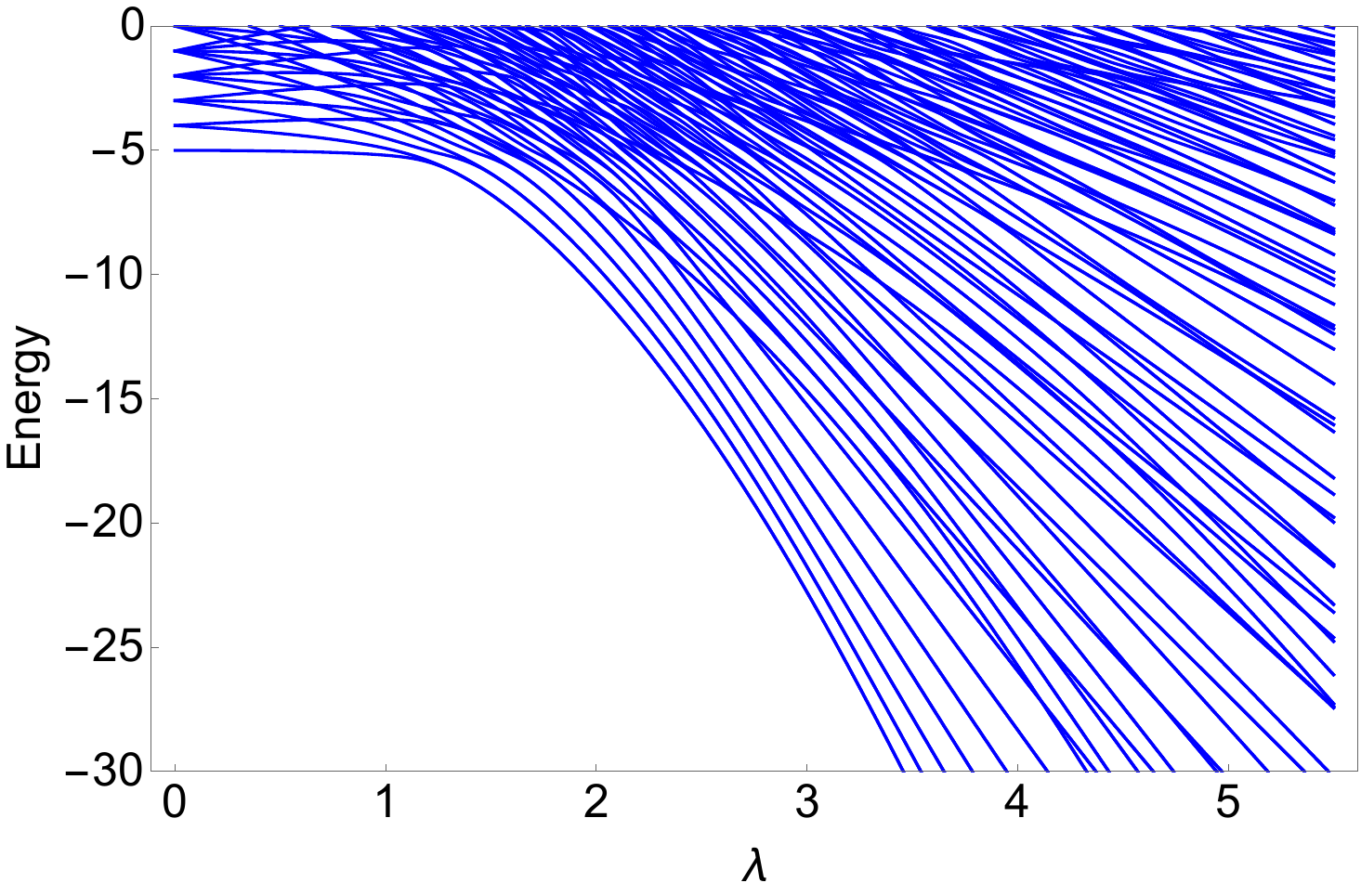}
        \caption{Equally balance between rotating and counter-rotating terms, $\gamma=0$.}
        \label{dickespec-a}
    \end{subfigure}
    \hfill
    \begin{subfigure}{0.49\textwidth}
        \centering
        \includegraphics[width=\linewidth]{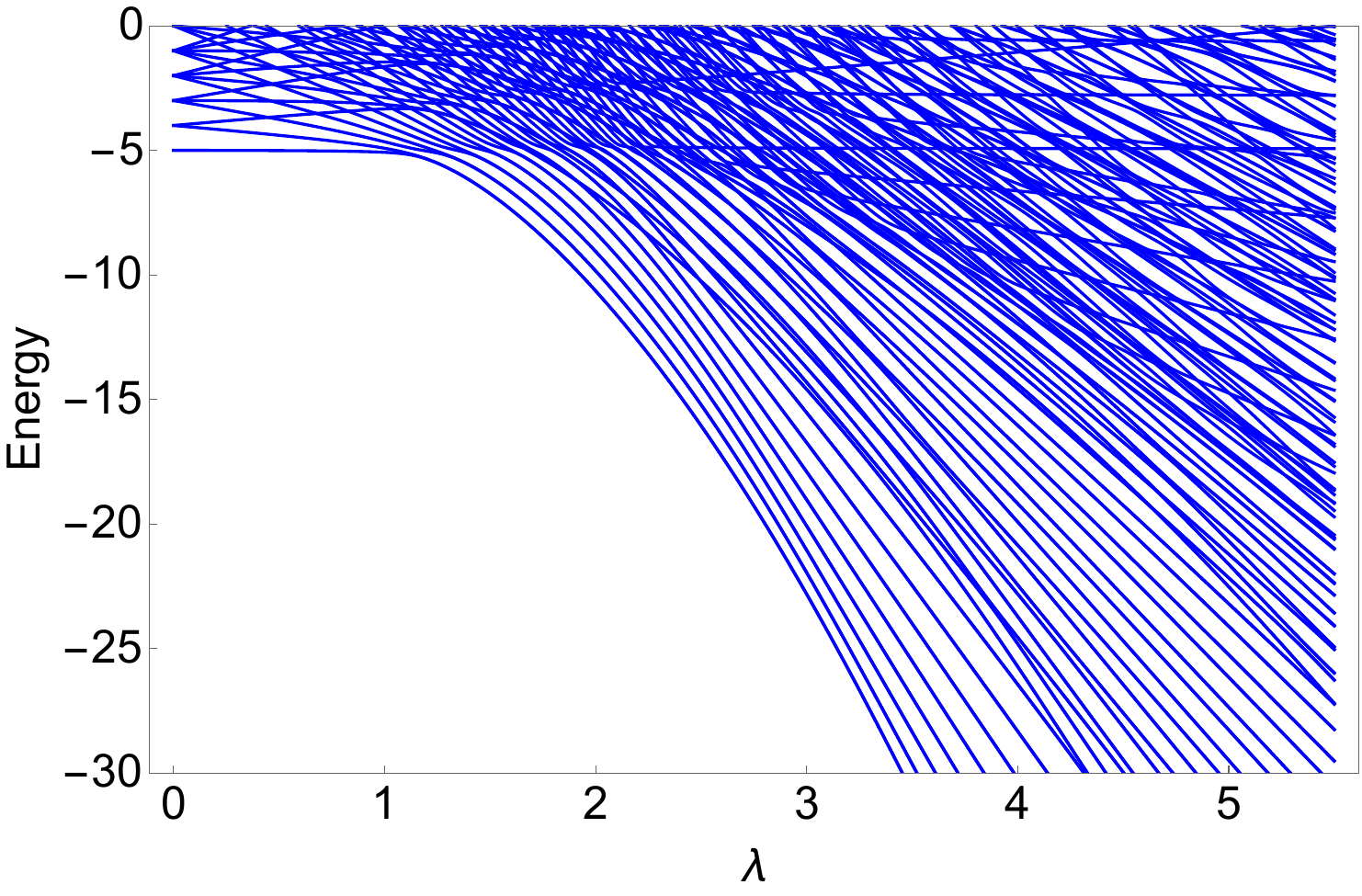}
        \caption{Preponderance of rotating wave terms over counter-rotating terms, $\gamma=0.5$.}
        \label{dickespec-b}
    \end{subfigure}    
    \begin{subfigure}{0.49\textwidth}
        \centering
        \includegraphics[width=\linewidth]{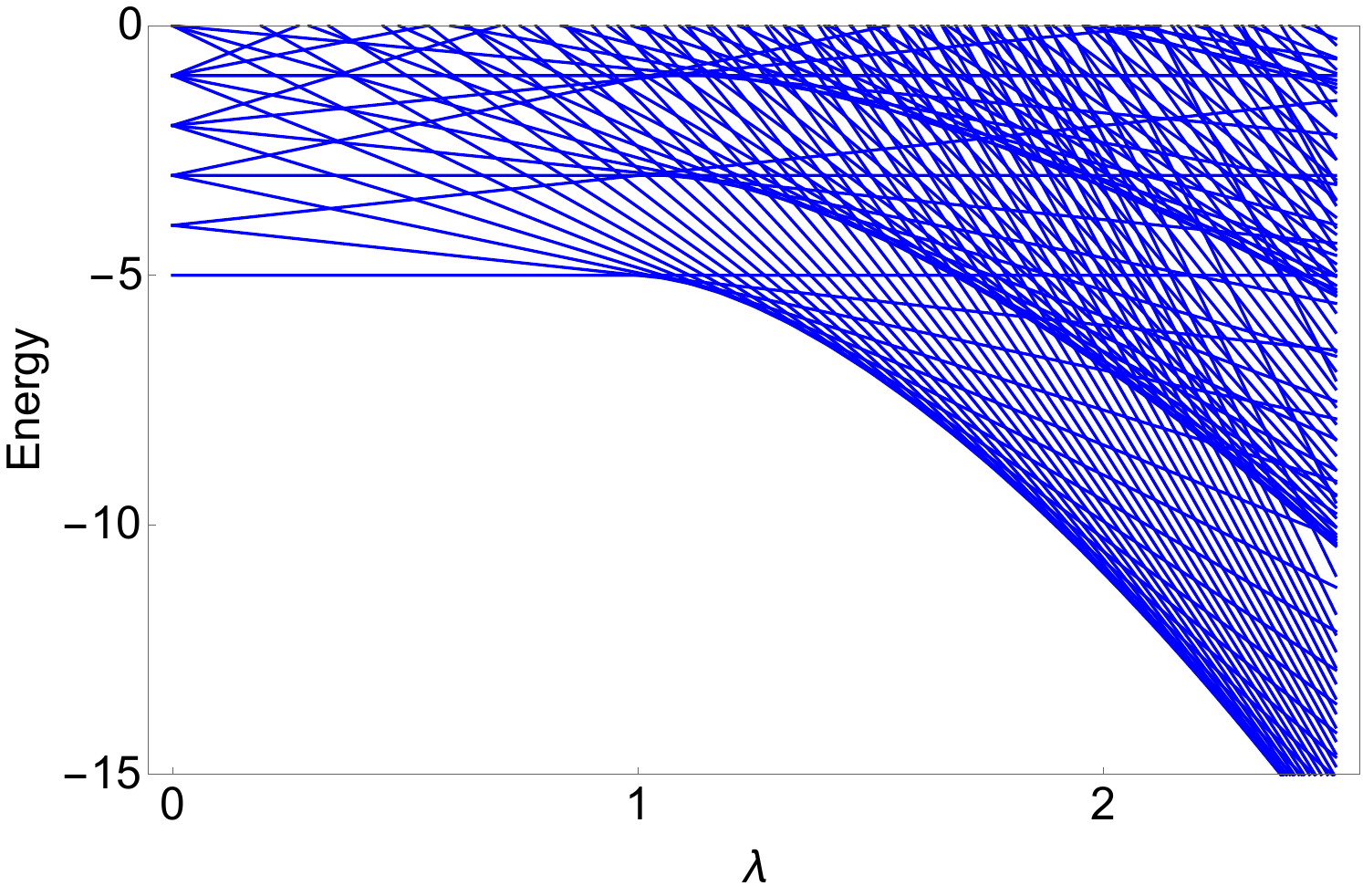}
        \caption{Rotating wave approximation  completely implemented, $\gamma=1$.}
        \label{dickespec-c}
    \end{subfigure}
    \caption{Spectrum of the full Dicke model as function of $\lambda$, for different values of $\gamma$. In all cases $\omega=\omega_0=1$, and the number of particles is $N=10$.}
    \label{figDickespec}
\end{figure}

The Dicke model is defined by an ensemble of two-level systems interacting with a common bosonic mode inside an optical cavity.
The quantum Hamiltonian is defined by
\begin{align}
{\cal H}_D&=\omega\,a^{\dagger}a+\omega_0J_z 
+\frac{\lambda(1+\gamma)}{2\sqrt{N}}\bigl(aJ_++a^{\dagger}J_-\bigr)
+\frac{\lambda(1-\gamma)}{2\sqrt{N}}\bigl(aJ_-+a^{\dagger}J_+\bigr)\,.
\label{HDicke}
\end{align}
The angular momentum operators are $J_{\alpha}\equiv\frac{1}{2}\sum_{i=1}^N\sigma_{\alpha}^{(i)}$, 
where $\sigma_\alpha^{(i)}$ are the usual Pauli matrices associated to the $i$-two level system, with $\alpha=x,y,z$; $J_{\pm}\equiv J_x\pm iJ_y$;  and the total angular momentum $(J_x,J_y,J_z)$ represents collective transitions in the set of $N$ two-level atoms. The annihilation and creation operators of the cavity bosonic mode are given by $a$ and $a^{\dagger}$, respectively. The energy gap between the two levels in each atom is $\omega_0$ and the energy of the bosonic mode is $\omega$.
The coupling constant between the bosonic mode and the two-level systems is $\lambda$
and the adimensional parameter $0\leq \gamma\leq1$  allows to weight 
the rotating and  the counter-rotating terms differently from each other. In this manner, $\gamma=0$ indicates the equally balance between rotating and counter-rotating terms, while $\gamma=1$ give us the rotating-wave approximation.

The model exhibits a second order quantum phase transition driven by $\lambda$ and
defined in the thermodynamic limit, i.e.  $N\rightarrow\infty$.
At zero temperature there is a transition from normal phase ($\lambda<\lambda_C=\sqrt{\omega\omega_0}$) to superradiant phase ($\lambda>\lambda_C=\sqrt{\omega\omega_0}$).
In order to analyze this, it is useful to define the number excitation operator  $\hat{\cal N}=a^{\dagger}a+J_z+N/2$ and the parity operator $\Pi=\ee^{\ii\pi\hat{\cal N}}$. It is possible to verify that the model possess the discrete parity symmetry, i.e. $[{\cal H},\Pi]=0$. In the special case where $\gamma=1$, rotating wave approximation, the model additionally possess the continuous symmetry $[\ee^{\ii\theta\hat{\cal N}},{\cal H}]=0$, for any $0\le\theta<2\pi$, since it that case $[{\cal H},\hat{\cal N}]=0$. The quantum phase transition occurs at the critical point where these respective symmetries are broken.

At finite values of $N$, the spectrum dependency on the coupling parameter $\lambda$ for the model is presented at  Fig. \ref{figDickespec}. 
It can be notice that a precursor of quantum phase transition is indicated by the breakdown of discrete parity symmetry  and  correspondingly the double degeneracy of the ground state
of the systems for the cases $\gamma\neq 1$. An important characteristic when $\gamma=1$, is the multiple level crossing for the lower energetic levels, in the superradiant phases. Such behavior is associated to the continuous symmetry breaking of the model in the cases where $\gamma=1$. Although the above results are valid only for the zero-temperature quantum phase transition, it can be shown that even at finite temperature the model exhibit a phase transitions identical to the quantum phase transition \cite{Aparicio2010, Aparicio2011}. 


\section{\label{sec:level3}Quantum 
Thermodynamics and Quantum Heat Engines}

\begin{figure}[b!]
	\begin{center}
		\includegraphics[width=.5\linewidth]{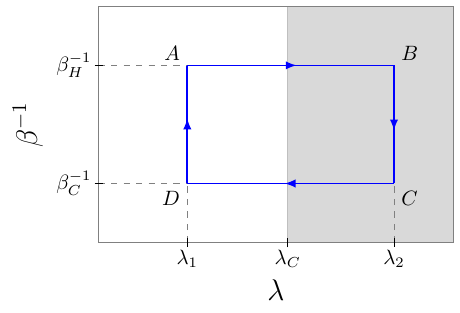}
	\end{center}
	\caption{Stirling cycle for the quantum heat engine studied, in a temperature $\times$ coupling constant phase diagram. The vertical line over $\lambda_C$ represents the quantum phase transition for the model, and the white and gray shadow regions represent respectively the normal and superradiant phases for the model.}
	\label{ciclo}
\end{figure}

In this section we  briefly review the notions of quantum thermodynamic. 
In the quantum domain the definitions of heat and work have to be revisited.
In order to accomplish this we can use the standard approach to distinguish the different kinds of energy exchanges of the quantum system.
Let us suppose that the system state is given by the 
density operator $\rho$ and that its dynamics is governed by the Hamiltonian ${\cal H}$. Hence, the system mean energy  is given by
$U=\tr\left({\rho {\cal H}}\right)$ and the energy exchange of the systems is
\begin{equation}
dU=\tr\left({d\rho\,{\cal H}}\right)+\tr\left({\rho\,d{\cal H}}\right)\,.
\label{1law}
\end{equation}
In Eq. (\ref{1law}), we identify the first term on the right hand side as the heat $\delta\langle Q\rangle$ absorbed by the system. 
This definition of heat is in agreement with the usual notion of heat as the energy variation associated with the change of the system internal state. This definition is compatible with the interpretation of heat as the energy variation related to an entropy increase. 
On the other hand, the second term in the right hand side of Eq. (\ref{1law}) is identified as the work performed over the system,
which is equivalent to the negative of the work realized by the system
$-\delta\langle W\rangle$. This definition relates work to the energy variation caused by changes in the external parameters that define the Hamiltonian of the system, as background fields or cavity size for example. It is worth mentioning that, in general work is process dependent and is not an observable, which probability distribution and characteristic function should be properly defined by a two-measurement protocol  \cite{Talkner2007}. Therefore, we have that the mean values of absorbed heat and work realized by the quantum system in a thermodynamic transformation are
\begin{align}
\langle Q\rangle&=\int\tr\left({d\rho\,{\cal H}}\right),\nn\\
\langle W\rangle&=-\int\tr\left({\rho\,d{\cal H}}\right),
\end{align}
these quantities are path dependent and their values depend on how the density state
or the Hamiltonian changes during the process. For our case, we assume that the system is always in 
thermal equilibrium with a thermal bath at temperature $\beta^{-1}$, and its density operator is a Gibbs state of the form
\begin{equation}
\rho(\beta,\lambda)=e^{-\beta{\cal H}(\lambda)}/{\cal Z}(\beta,\lambda),
\label{gibbs}
\end{equation}
where ${\cal Z}(\beta,\lambda)=\tr(e^{-\beta{\cal H}(\lambda)})$ is the partition function and ${\cal H}(\lambda)$ is the system Hamiltonian parametrized by a general external variable $\lambda$. In our study, we consider a Hamiltonian with the coupling constant $\lambda$ which drives the quantum phase transition in the model.
We also consider a quantum Stirling heat engine, defined by two isothermals (i. e. constant $\beta$) and two processes with fixed value of coupling constant $\lambda$, see Fig. \ref{ciclo}.
For these processes we are able to rewrite our formal expressions of  mean quantum heat and  mean quantum work.
By using the notions of entropy and free energy
\begin{align}
S&=-\tr\left(\rho\ln\rho\right),\nn\\
F&=-\ln{\cal Z}/\beta,
\end{align}
one can show that if we are always on a Gibbs state Eq. (\ref{gibbs}) and the processes are such that probabilities are conserved, $\tr(d\rho)=0$, 
we have that
\begin{align}
\langle Q\rangle&=\int\frac{dS(\beta,\lambda)}{\beta},\nn\\
\langle W\rangle&=-\int\frac{\partial F(\beta,\lambda)}{\partial\lambda}d\lambda,
\label{QW}
\end{align}
here we need to put into evidence that the entropy and free energy depend on both temperature $\beta^{-1}$ and the coupling constant $\lambda$. 
Hence, in general, the amount of heat absorbed and work realized are path dependent.
Nonetheless, the first law of thermodynamics is always obeyed, since for any infinitesimal process we have
$dU=\delta\langle Q\rangle-\delta\langle W\rangle$. Therefore, in a finite process we have
\begin{equation} 
\Delta U=\langle Q\rangle-\langle W\rangle.
\label{1lawb}
\end{equation}
For the Stirling quantum heat engine we have two kind of processes:\\

\noindent
{\bf \em  Isothermal transformation}: In this process the temperature $\beta^{-1}$ of the system is kept constant by exchanging heat with a reservoir, and it realizes work over the environment. In this transformation the parameter $\lambda$ changes from some initial value $\lambda_i$ to a final one $\lambda_f$.
Hence, from Eq. (\ref{QW}) we have that in an isothermal process, the heat and work respectively read

\begin{align}
\langle Q\rangle_{isoth}&=\Delta S/\beta=(S(\beta,\lambda_f)-S(\beta,\lambda_i))/\beta,\nn\\
\langle W\rangle_{isoth}&=-\Delta F=-F(\beta,\lambda_f)+F(\beta,\lambda_i).
\label{}
\end{align}

\noindent
{\bf \em $\lambda$-fixed transformation}: In this process the external parameter $\lambda$, that defines the Hamiltonian, is kept constant while the temperature of the system $\beta^{-1}$ changes. Through the process, the system does not realizes any work, but it could exchange heat with the reservoir.
Since in this case $\lambda$ remains constant and the temperature changes from some initial value $\beta^{-1}_i$ to a final one $\beta^{-1}_f$, we have from Eqs. (\ref{QW}) and (\ref{1lawb}) that in the $\lambda$-fixed process
\begin{align}
\langle Q\rangle_{\lambda-fixed}&=U(\beta_f,\lambda)-U(\beta_i,\lambda),\nn\\
\langle W\rangle_{\lambda-fixed}&=0.
\label{}
\end{align}

As we just saw above, the thermodynamic processes depend on the internal energy $U$, entropy $S$ and free energy $F$, where they can be obtained from the partition function ${\cal Z}$. 
The computation of this partition function depends on the Hamiltonian and the Hilbert space of the system.
Moreover, the thermodynamic cycle of our quantum heat engine is the Stirling cycle, it is shown in Fig. \ref{ciclo} and defined by the following processes:\\

%

\noindent
{\bf \em Process $A\rightarrow B$:} Isothermal process at fixed temperature $\beta_H^{-1}$. In this process the coupling constant 
$\lambda$ changes from $\lambda_{1}$ to $\lambda_{2}$, going through the critical value $\lambda_C$ when $\lambda_{2}>\lambda_C$. Here the absorbed heat of the
system is given by
\begin{align}
\langle Q\rangle_{AB}=(S_B-S_A)/\beta_H>0\,.
\label{qab}
\end{align}
{\bf \em Process $B\rightarrow C$:} Process at fixed coupling constant $\lambda_{2}$.
Here, the temperature of the system diminishes from $\beta_H^{-1}$ to $\beta_C^{-1}$. The system does not realize work but releases heat 
given by 
\begin{align}
\langle Q\rangle_{BC}=U_C-U_B<0\,.
\label{qbc}
\end{align}
{\bf \em Process $C\rightarrow D$:} Isothermal process at fixed temperature $\beta_C^{-1}$. In this transformation the coupling constant $\lambda$ goes back from $\lambda_{2}$ to its initial value $\lambda_{1}$, going through the critical value $\lambda_C$ when $\lambda_{2}>\lambda_C$. In this process it releases heat given by
\begin{align}
\langle Q\rangle_{CD}=(S_D-S_C)/\beta_C<0\,.
\label{qcd}
\end{align}
{\bf \em Process $D\rightarrow A$:} Process at fixed coupling constant $\lambda_{1}$.
Here the temperature increases from $\beta_C^{-1}$ to $\beta_H^{-1}$. The system does not realize work but absorbs heat 
given by
\begin{align}
\langle Q\rangle_{DA}=U_A-U_D>0\,.
\label{qda}
\end{align}
For the whole cycle, the efficiency of the quantum Stirling heat engine is given by
\begin{align}
\eta =\frac{\langle W\rangle}{\,\,\,\,\,\langle Q\rangle_{abs}}\,,
\label{efic1}
\end{align}
where $\langle W\rangle$ is the total work performed by the system and 
$\langle Q\rangle_{abs}=\langle Q\rangle_{AB}+\langle Q\rangle_{DA}$, is the amount of heat absorbed by the system. Since we have a cyclic process, and due to the first law of thermodynamics the total change in energy is zero, we have $\langle W\rangle=\langle Q\rangle_{AB}+\langle Q\rangle_{BC}+\langle Q\rangle_{CD}+\langle Q\rangle_{DA}$.

\section{\label{sec:level4} Efficiency in the Dicke model Quantum  Heat Engine }

\begin{figure}[h!]
    \centering
    \begin{subfigure}{0.49\textwidth}
        \centering        \includegraphics[width=\linewidth]{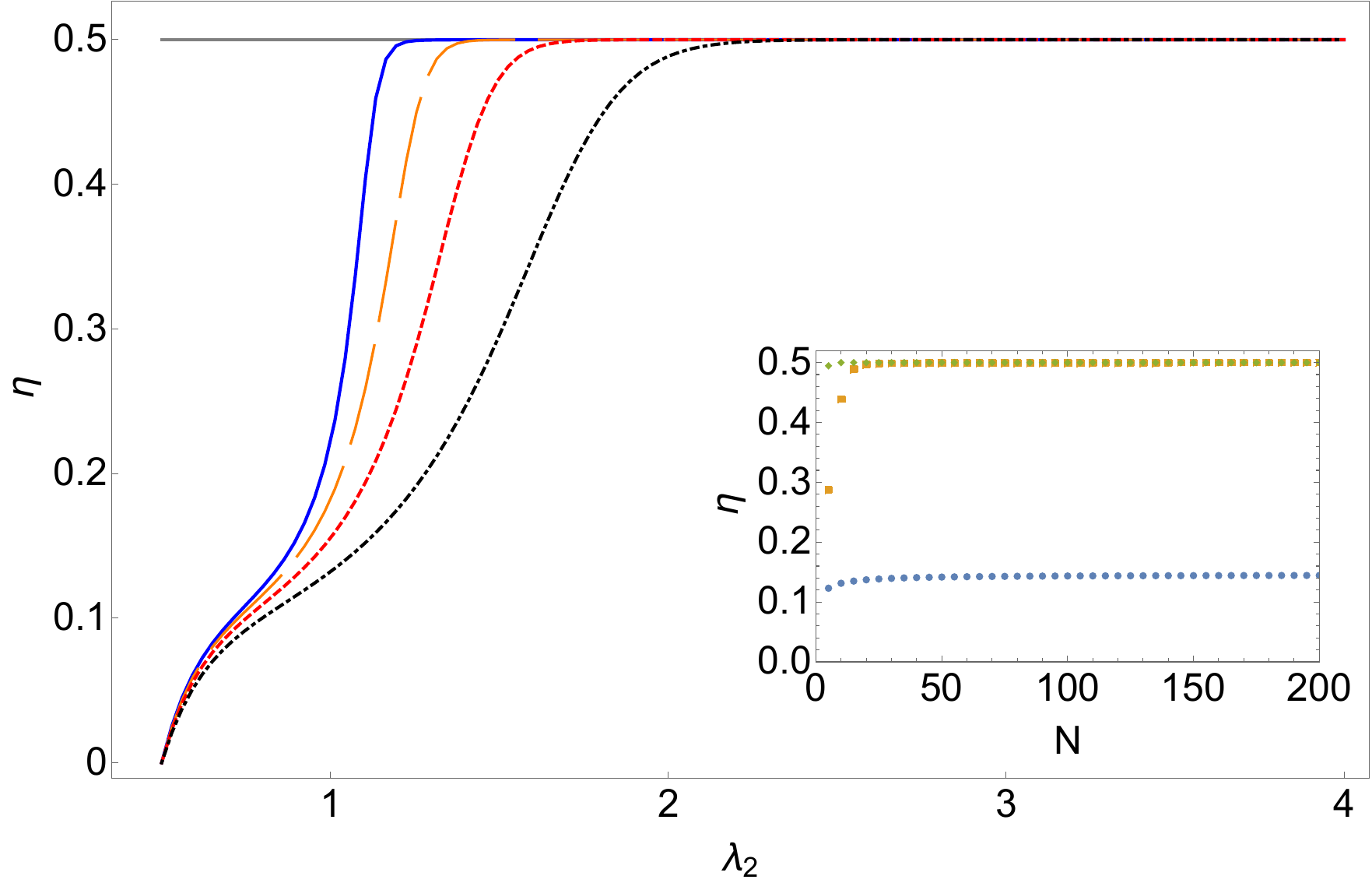}
        \caption{Equally balance case, $\gamma=0$.}
    \end{subfigure}
    \hfill
    \begin{subfigure}{0.49\textwidth}
        \centering        \includegraphics[width=\linewidth]{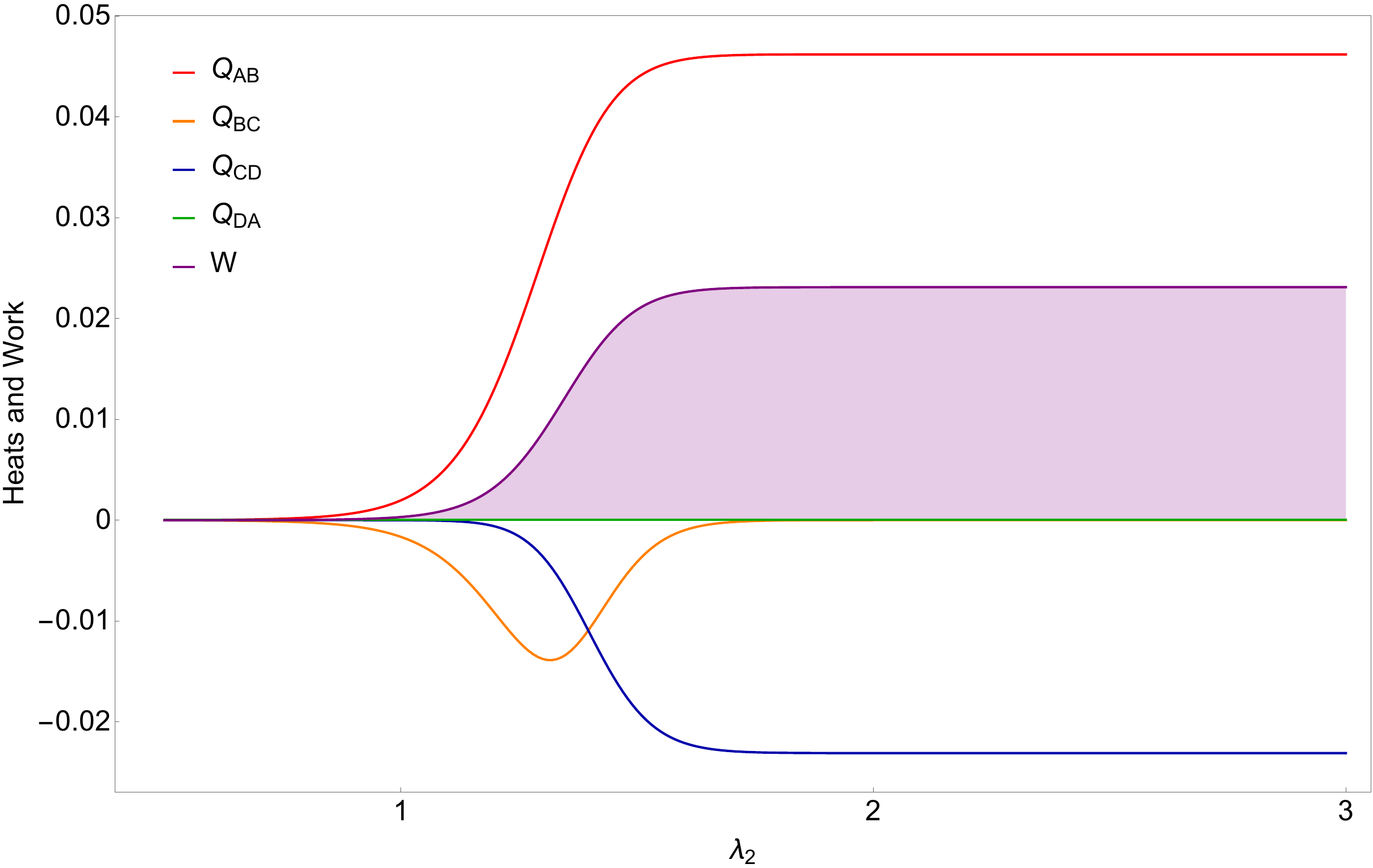}
        \caption{Equally balance case, $\gamma=0$.}
    \end{subfigure}    
    
    \vspace{10pt}
    \begin{subfigure}{0.49\textwidth}
        \centering        \includegraphics[width=\linewidth]{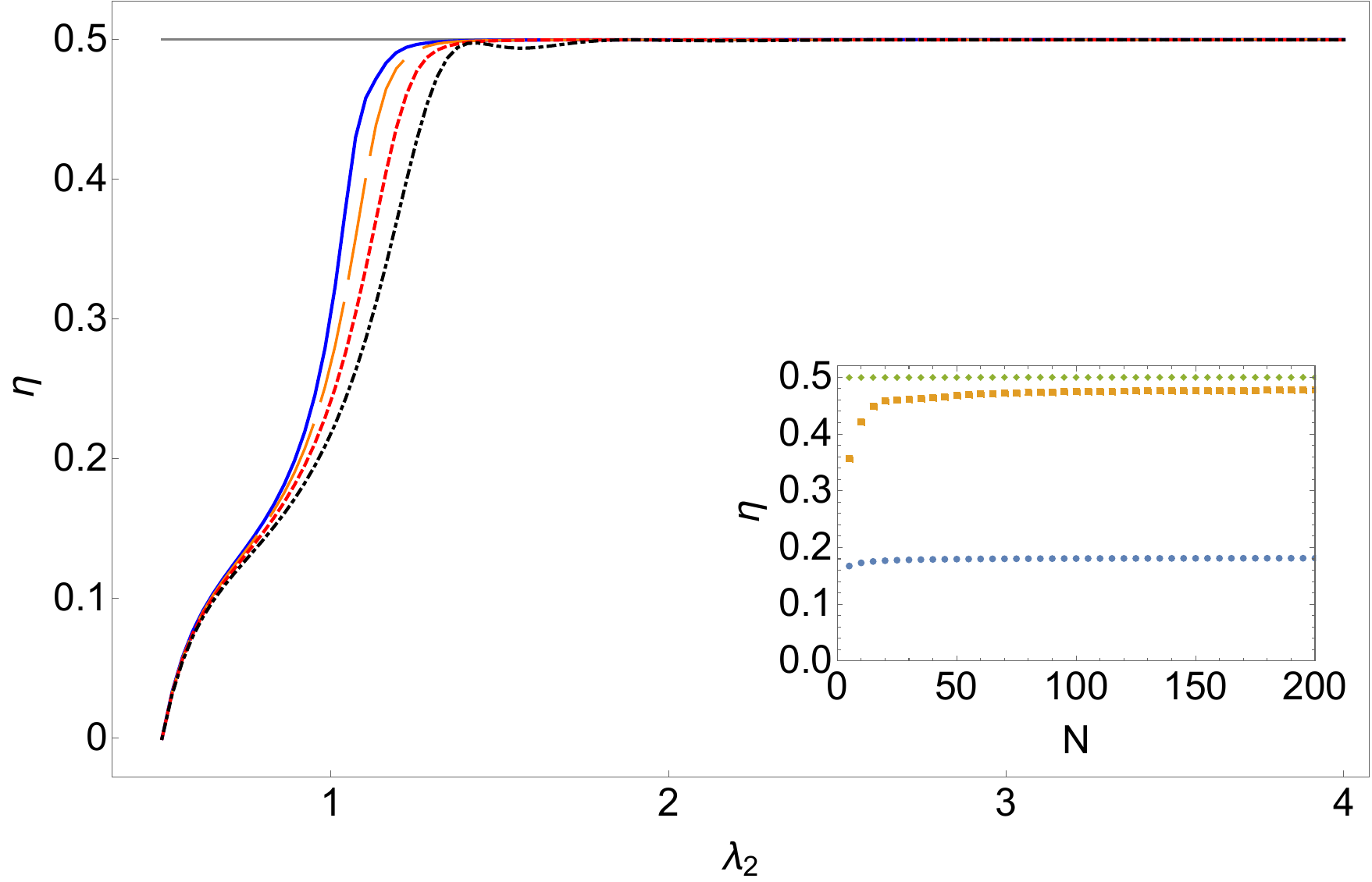}
        \caption{Preponderance of rotating terms over counter-rotating terms, $\gamma=0.5$.}
    \end{subfigure}
    \hfill
    \begin{subfigure}{0.49\textwidth}
        \centering        \includegraphics[width=\linewidth]{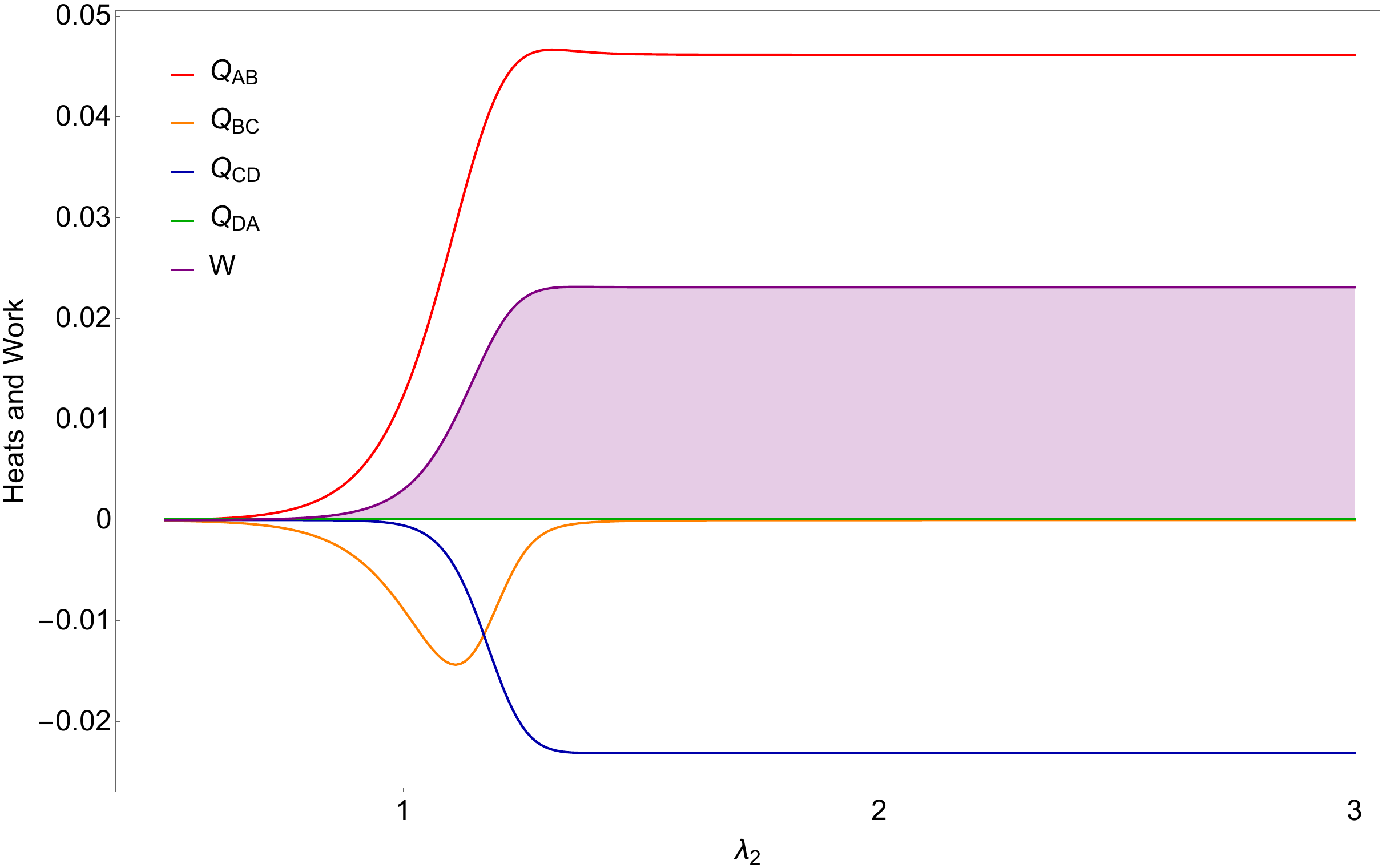}
        \caption{Preponderance of rotating terms over counter-rotating terms, $\gamma=0.5$.}
    \end{subfigure}
    \vspace{10pt}
    
    \begin{subfigure}{0.49\textwidth}
        \centering        \includegraphics[width=\linewidth]{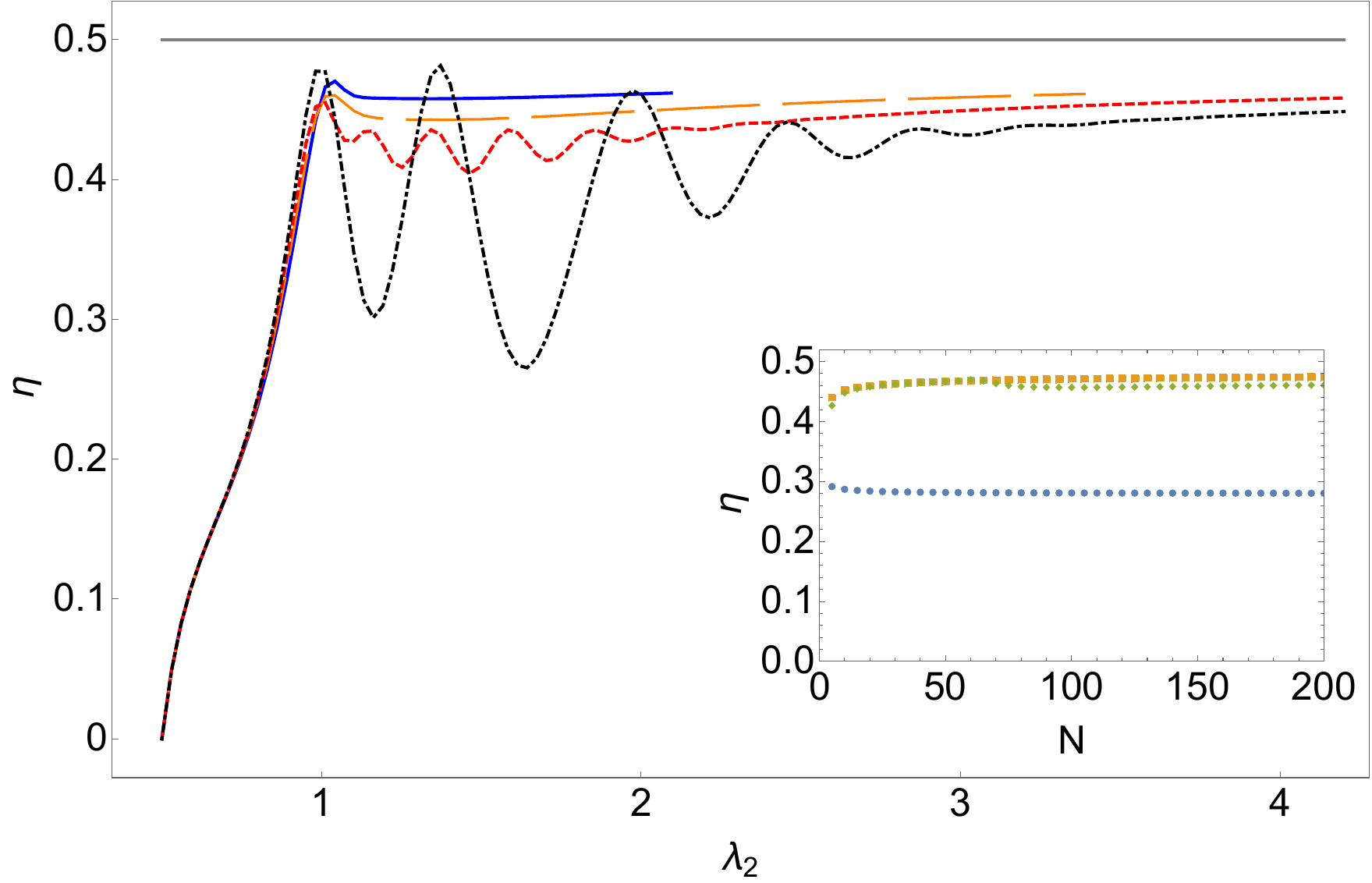}
        \caption{Rotating wave approximation, $\gamma=1$.}
    \end{subfigure}
    \hfill
    \begin{subfigure}{0.49\textwidth}
        \centering        \includegraphics[width=\linewidth]{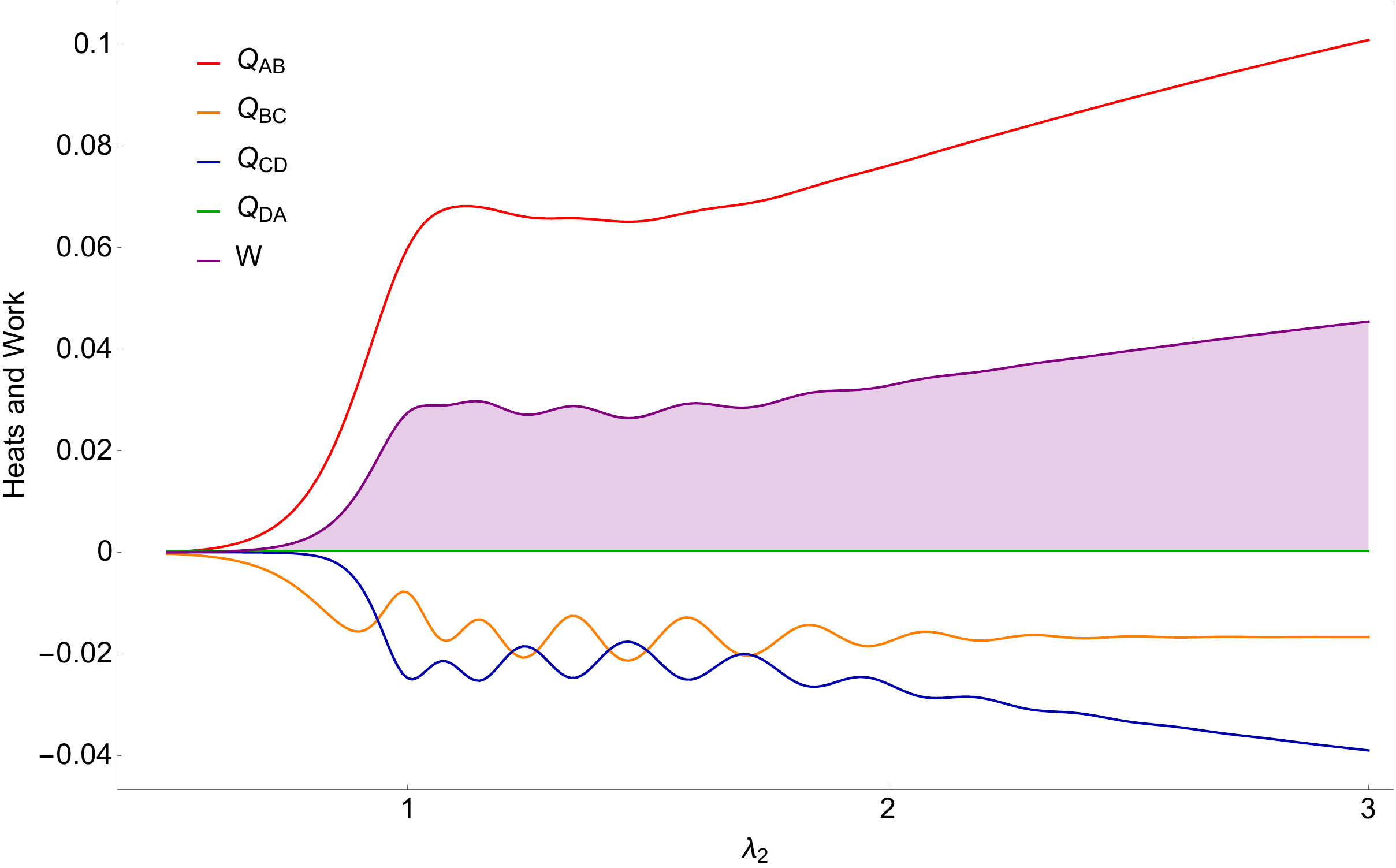}
        \caption{Rotating wave approximation, $\gamma=1$}
    \end{subfigure}    
    \vspace{10pt}
    \caption{The quantum heat engine efficiency for the Dicke model as function of $\lambda_2$ is shown in panels (a), (c) and (e) for different values of $\gamma$ and for number of particles $N=20$ (blue full line), $N=8$ (orange dashed line), $N=4$ (red dot-dashed line) and $N=2$ (black dotted line). The gray upper line correspond to the Carnot efficiency. The heat exchanged at each path of the cyclic process, see Fig. \ref{ciclo}, as function of $\lambda_2$ are shown in panels (b), (d) and (f) for different values of $\gamma$ and with $N=4$. For all the panels: $\omega=\omega_0=1$, $\beta_C^{-1}=1/30$, $\beta_H^{-1}=1/15$ and $\lambda_{1}=0.5$.}
    \label{figdicke}
\end{figure}

\begin{figure}[t!]
    \centering
        \begin{subfigure}{0.6\textwidth}
        \centering
        \includegraphics[width=\linewidth]{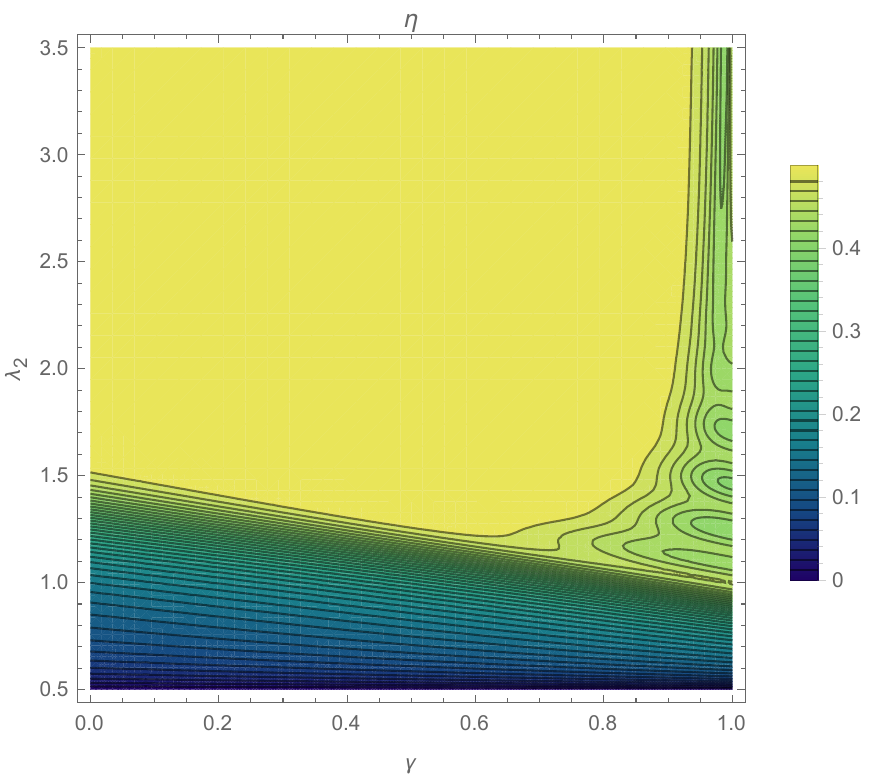}
    \end{subfigure}        
    \vspace{10pt}
    \caption{Efficiency of the full Dicke model quantum heat engine as function of the parameter $\gamma$ and the coupling constant $\lambda_2$. Regions of higher efficiency are found for smaller values of $\gamma$ (equally balance between rotating and counter-rotating terms), and higher values of $\lambda_2$. In the regions of greater $\gamma$, more regions of lower efficiency appear. We used $\omega=\omega_0=1$, $\beta_C^{-1}=15$, $\beta_H^{-1}=30$, $\lambda_1=0.5$ and $N=4$.}
    \label{gammalambda}
\end{figure}

In this section we show numerical results for the work and efficiency of the quantum heat engine defined in the last section. The system is always at thermal equilibrium, then its state is the Gibbs state, see Eq. (\ref{gibbs}). In such context, the thermodynamic quantities are obtained from the partition function, which is computed by numerical diagonalization of the Hamiltonian, given by Eq.  (\ref{HDicke}). This technique requires a cut-off in the infinite Hilbert space of the bosonic mode. The cut-off was set up demanding a good convergence of the numerical results.
The principal results  are shown in Fig.  \ref{figdicke} and Fig.  \ref{gammalambda}. 
In the calculations we set the atoms energy gap  and the bosonic energy excitations to be $\omega_0=\omega=1$, consequently the quantum phase transition occurs at $\lambda_C=1$.
We fix the value $\lambda_1=0.5$ to ensure that part of the thermodynamic cycle (sates $A$ and $D$ in Fig. \ref{ciclo}) is necessarily on the normal phase and allow $\lambda_2$ to change.
It is observed, from Fig. \ref{figdicke}, that the efficiency has lower values when $\lambda_2$ satisfies $\lambda_1<\lambda_2<\lambda_C$, i.e. the whole cycle remains in the normal phase of the full Dicke model.

When we allow the system to undergoes  a quantum phase transition during the thermodynamic cycle, in such a way that points $B$ and $C$ of the cycle are now in  the super-radiant phase, i.e. $\lambda_2>\lambda_C$, we obtain a great enhancement of the quantum heat engine efficiency. Moreover, the cycle efficiency eventually reaches the maximum efficiency dictated by the Carnot limit.

The effect over the cycle efficiency due to 
the unbalance of rotating and counter-rotating terms, meaning 
 different values of $\gamma$, are also shown in Fig. (\ref{figdicke}).
In this sense, in Fig. (\ref{figdicke}a)
 we have the equally balance case 
 setting $\gamma=0$. It can be seen that the efficiency converges to the Carnot efficiency after $\lambda_{2}$ exceeds the critical value $\lambda_C$. We also notice that this convergence to maximum efficiency is more rapidly obtained for the perfectly balance situation ($\gamma=0$), when compared to cases where the rotating wave approximation is dominant ($\gamma\neq0$). 
The case with $\gamma=0.5$ is shown in Fig. (\ref{figdicke}c), while the case $\gamma=1$ related to the Dicke model with rotating wave approximation is presented in Fig. (\ref{figdicke}e). It can be notice from these figures that an oscillatory behavior of efficiency appears, as $\lambda_2$ is increased, and we can show that these oscillations are more prominent for the case $\gamma\approx 1$. Hence, a preponderance of rotating terms in the full Dicke model prevents to reach the maximum efficiency. The quick convergence of the quantum heat engine efficiency towards its Carnot bound when a balance between rotating and counter-rotating terms is consider, is an indication that the rotating wave approximation can be harmful to the quantum heat engine performance.

In order to have a broader visualization of the effect of unbalance between rotating and counter-rotating terms in the full Dicke model quantum heat engine we present Fig. \ref{gammalambda}.
This graph shows the behavior of the quantum heat engine efficiency for different values of $\gamma$ and coupling constant $\lambda_2$, simultaneously. 
It can be notice that for small values $\gamma\approx 0$ (equally balance between rotating and counter-rotating terms), the efficiency rapidly grows up to a maximum value, specially after $\lambda_2$ overpass $\lambda_C$ approximately. In the case of $\gamma\approx 1$, regions of small efficiencies appear and the efficiency oscillates as we increase the values of $\lambda_2$ for fixed $\gamma$.


In order to inspect the converge towards the maximum
quantum heat engine efficiency, given by the Carnot bound, we also show in Fig. \ref{figdicke} the behavior of the heat exchange in each part of the cycle as well as the total work of the cycle. The energies associated with the cycle for different values of $\gamma$ are presented at Figs. (\ref{figdicke}b, \ref{figdicke}d, \ref{figdicke}e). Since the quantum heat engine operates on a Stirling cycle, the heat $Q_{AB}$ represents the heat absorbed during the isothermal process at high temperature $\beta_H^{-1}$ that change the state of the system from a normal phase $\lambda_1$ to some other phase $\lambda_2$.
The heat $Q_{BC}$ is the heat released from the system to diminish its temperature from $\beta_H^{-1}$ to $\beta_C^{-1}$ while staying constant at the phase $\lambda_2$. The heat $Q_{CD}$ is the heat extracted from the system to return from the phase $\lambda_2$ to the normal phase $\lambda_1$ at a constant lower temperature $\beta_C^{-1}$. Finally, $Q_{DA}$ is the heat absorbed by the system to increase its temperature from $\beta_C^{-1}$ to $\beta_H^{-1}$ while staying at the normal phase $\lambda_1$.
From Figs. (\ref{figdicke}b, \ref{figdicke}d, \ref{figdicke}e) it can be notice that the heat $Q_{DA}$ is positive but very small when compared with the other heats. Then this process, $DA$, can be consider approximately as an adiabatic process. On the other hand, the heat $Q_{BC}$ is always negative, reaching some maximum negative value near the phase transition, and then tends to zero. Hence, after the critical point the process $BC$ can be also considered as an adiabatic process with no heat exchange. In this final situation we have that the Stirling cycle would effectively consist of a pair of adiabatic process ($BC$ and $DA$) and a pair of isothermal process ($AB$ and $CD$) leading us to effectively a Carnot cycle.

\section{\label{sec:level5}{Conclusions ans Perspectives}}

The efficiency of a quantum heat engine was computed for an engine operating with a quantum critical system, specifically the full Dicke model, where the thermodynamic cycle is of the Stirling type.  For the model, we show numerical results for low temperatures, where optimization of the efficiency value (reaching the Carnot limit at some particular cases) was observed, it happens when during the cycle process the model undergoes a quantum phase transition. Since quantum phase transition happens for $N\rightarrow\infty$, in the numerical study, some scaling graphs were performed, which show the behavior of efficiency near to the thermodynamical limit ($N\rightarrow\infty$). The efficiency, and other thermodynamical quantities related to the quantum heat engine, for finite $N$ also were studied.
A faster increase in efficiency is achieved when the interaction term in the  interaction posses the rotating and counter-rotating terms equally balanced. This means that the rotating wave approximation (i.e. when the counter-rotating term is eliminated)  is not suitable for achieving a faster increase in efficiency.

It is observed that Carnot and our Stirling cycles are equivalent for certain values of the model parameters, where in such cases, there are heat transference only in the isothermal processes of the Stirling cycles. In that cases efficiency maximum values are attained and it is also noticed how the ground state degeneracies (related to the quantum phase transition) are associated to such maximum efficiencies. Such behavior is related to symmetries of the model, since symmetries rule the degeneracy of the ground state due to the quantum phase transition and also rule the existence of level crossings in the spectrum of the systems.

Is important to mention that the thermodynamics of the models studied here, correspond to indistinguishable particles in a two level system. For this situation, the Hilbert space of the set of $N$ particles is $N-$dimensional, and thermodynamic quantities are not extensive. 
In the last case thermal phase transitions are observed, nevertheless in the indistinguishable cases do not, where only appears quantum phase transition (see \cite{Dicke05} for the Dicke model in the ultrastrong-coupling regime). An interesting extension of this work  is to study 
 the Lipkin-Meshkov-Glick model that describes an ensemble of two-level system with all-to-all interaction. The Lipkin-Meshkov-Glick model exhibits a transition from paramagnetic phase ($\lambda<\lambda_C=\omega_0$) to ferromagnetic phase ($\lambda>\lambda_C=\omega_0$). A recent theoretical study \cite{ma} defines a quantum heat engine with the isotropic Lipkin-Meshkov-Glick model. The authors proved that higher values of efficiency are obtained (getting Carnot's value at some cases) when during the thermodynamic cycle the substance undergoes a quantum phase transition. The study of the anisotropy in the self-interaction in the Lipkin-Meshkov-Glick model and its relation to the quantum heat engine efficiency is under investigation by the authors.

\section*{Acknowledgement}
This work was partially supported by Conselho Nacional de Desenvolvimento Cient\'{\i}fico e Tecnol\'{o}gico - CNPq, the grant - 305000/2023-3 (N.F.S).
E. A. wish to acknowledge the support of FAPERJ, Fundação Carlos Chagas Filho de Amparo à Pesquisa do Estado do Rio de Janeiro, 
Process Number E-26/010.101230/2018.

\appendix

\section{\label{apdeg}Ground state double degeneracy and efficiency increase in quantum heat engine}
\begin{figure}[h!]
\centering
	\includegraphics[width=0.55\linewidth]{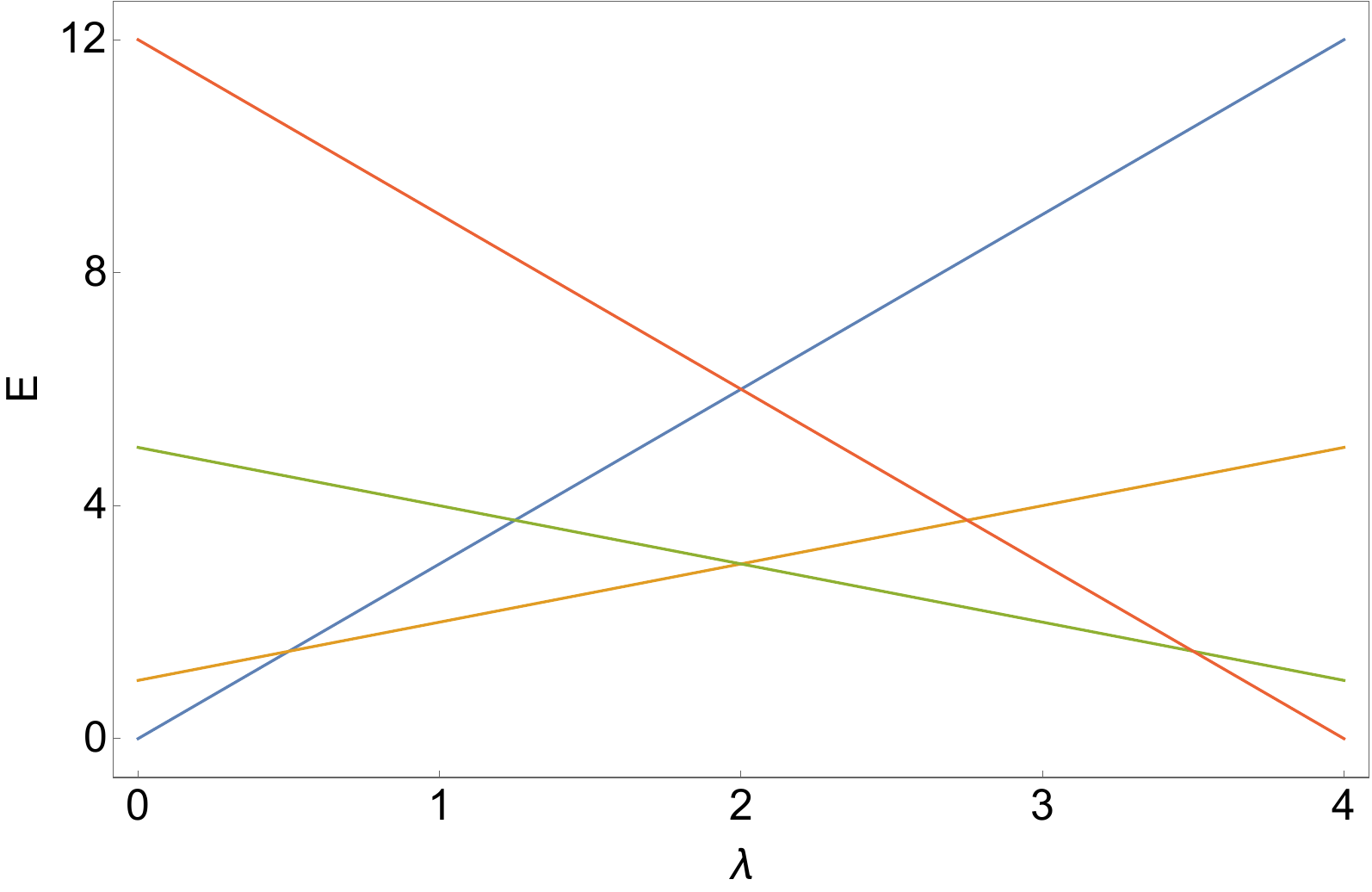}
	\caption{Crossing energy levels of a toy system. Each curve and respective energies function are: blue for $E_1(\lambda)=3\lambda$, yellow for $E_2(\lambda)=1+\lambda$, green for $E_3(\lambda)=5-\lambda$ and red for $E_4(\lambda)=12-3\lambda$.}
	\label{cruces}
\end{figure}

In this appendix we closely examine the relationship between the efficiency increase of the quantum heat engines
and the appearance of a double degeneracy in the ground energy level during the process. It is known that for the models studied in this work, such behavior happens at the quantum phase transition. The appearance of a double degeneracy is observed also in Figs. \ref{figDickespec}. In order to do the examination aforementioned, we analyze a simpler system (toy system). This system possess 4 energy levels dependent on some external parameter $\lambda$, in such a way that by modifying its value we are able to control the existence of the level crossings. Let us set the four energy levels given by $E_1(\lambda)=3\lambda$, $E_2(\lambda)=1+\lambda$, $E_3(\lambda)=5-\lambda$ and $E_4(\lambda)=12-3\lambda$; plotted in Fig.(\ref{cruces}).
There we can see that for values of $\lambda$ lower that $0.5$, the ground state corresponds to the first energy level $E_1(\lambda)$, but as $\lambda$ is increased the ground state changes $|E_1\rangle\rightarrow|E_2\rangle\rightarrow|E_3\rangle\rightarrow|E_4\rangle$.
We also identify three values of $\lambda$ where the lower energetic level crossings happen, at
$\lambda=\{0.5, 2, 3.5\}\,.$
Similar as our study of the full Dicke model, we calculate the partition function for this toy system assuming thermal equilibrium with a reservoir at temperature $\beta^{-1}$, it means, the partition function is: ${\cal Z}(\beta,\lambda)=\sum_{i=1}^{i=4}e^{-\beta\,E_i(\lambda)}$. Moreover, the quantum heat engine cycle is of the Stirling type, as defined in section \ref{sec:level3}, see Fig. \ref{ciclo}. 

\begin{figure}[t!]
\centering
	\includegraphics[width=0.55\linewidth]{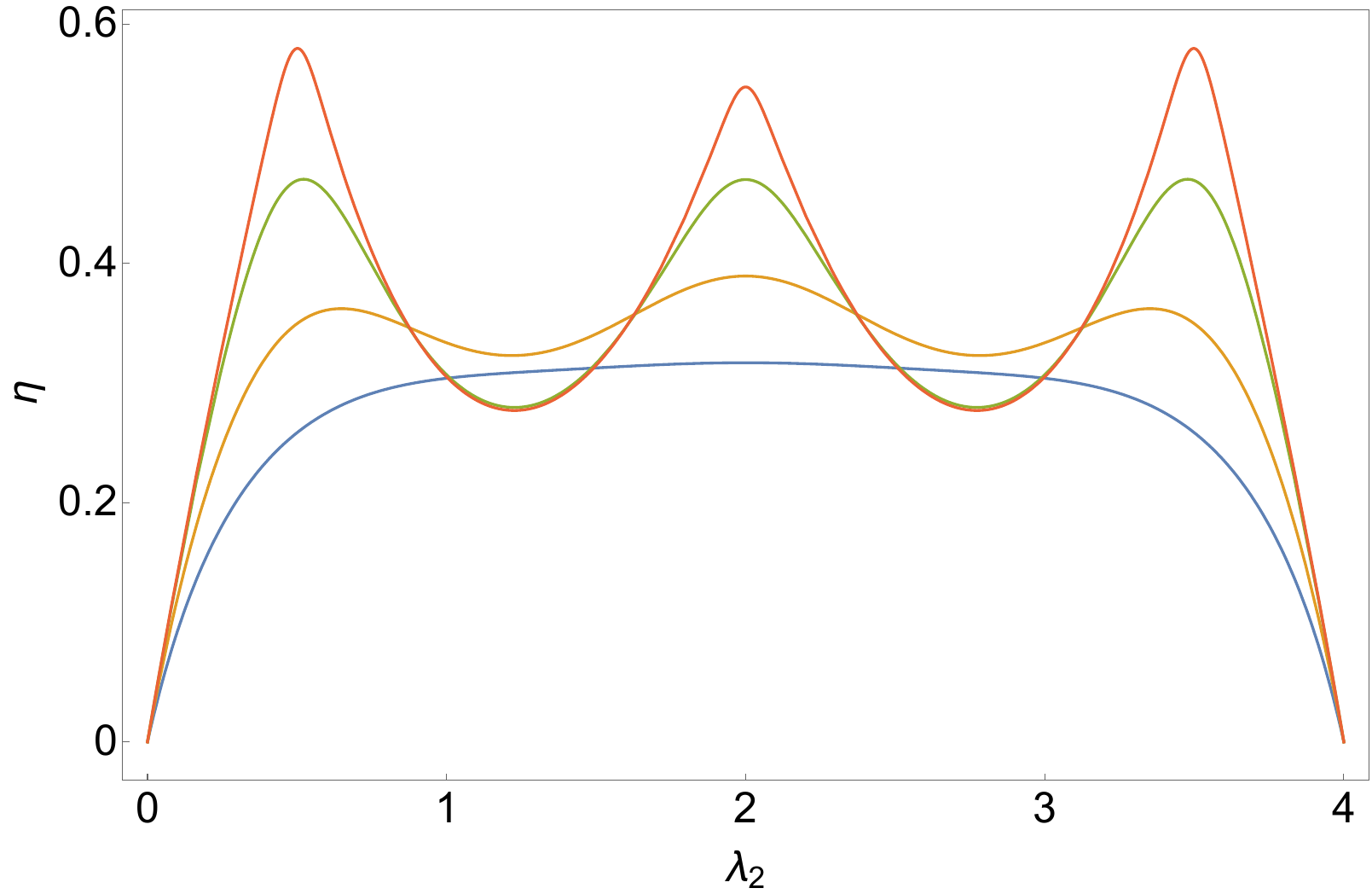}
	\caption{Graphs of efficiency function for the toy system at thermal equilibrium in an Stirling cycle (see Fig. \ref{ciclo}). In whole the curves: $\lambda_1=0$, hot reservoir temperature $\beta_H^{-1}=2$ and $\lambda_2$ varies from $0$ to $4$. Temperature values of the cold reservoir are $\beta_C^{-1}=1$ for the blue curve, $\beta_C^{-1}=0.5$ for the yellow curve, $\beta_C^{-1}=0.2$ for the green curve and $\beta_C^{-1}=0.05$ for the red curve.}
	\label{efftoy}
\end{figure}

\vspace{1cm}
Results for the quantum heat engine efficiency  are shown in Fig. \ref{efftoy}, where some cycle parameter values are $\lambda_1=0$ and $\lambda_2$ varies from $0$ to $4$. So when $\lambda_2$ goes through a level crossings (at values $\lambda_2=\{0.5, 2, 3.5\}$) an increase of efficiency happens. It is noticed also that such increase is more accentuated at low temperatures.

\begin{figure}[t!]
\centering
	\includegraphics[width=0.55\linewidth]{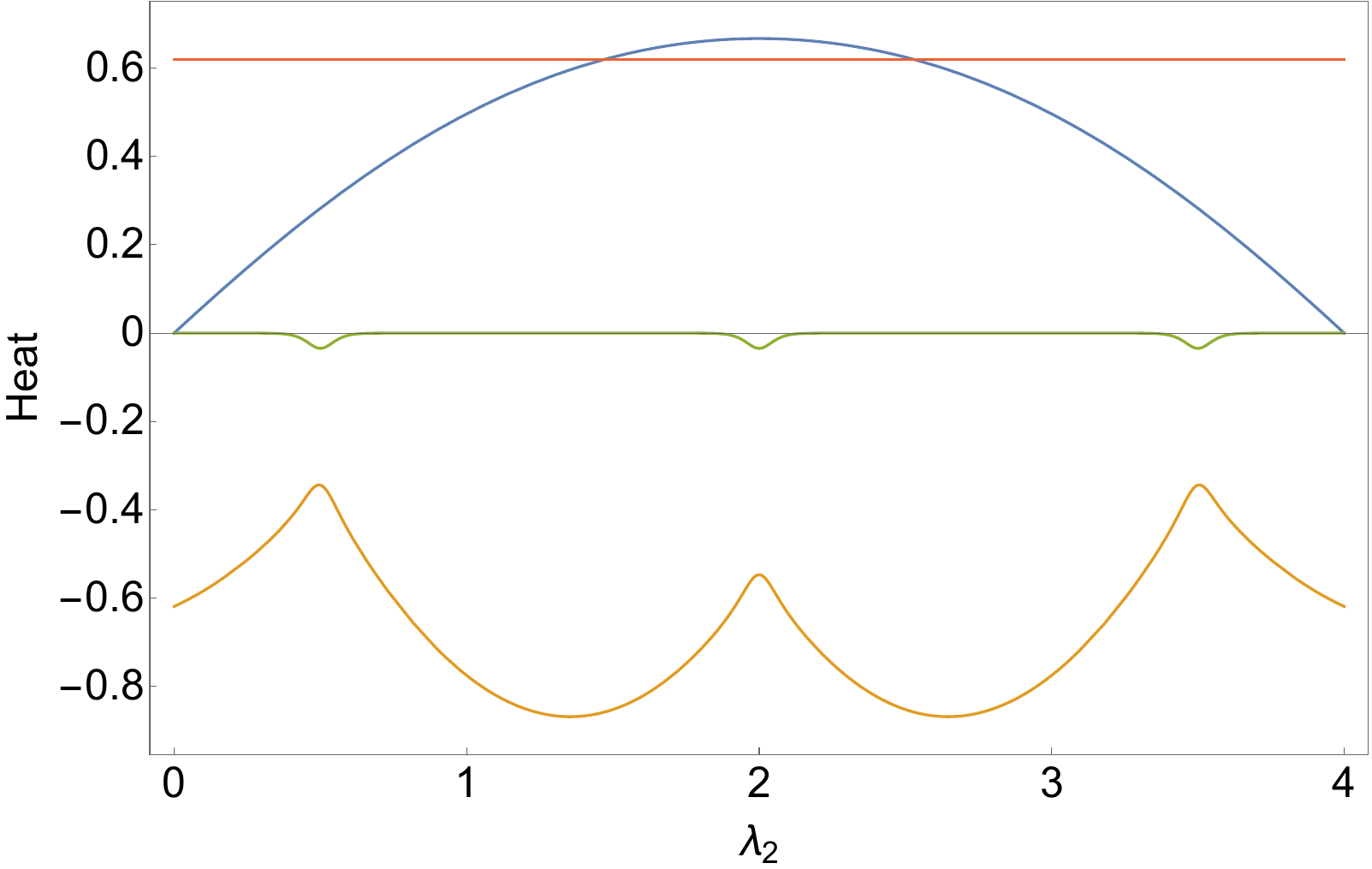}
	\caption{Heats involved in the quantum heat engine process for the toy system, according to the Stirling cycle (see Fig. \ref{ciclo}). Was set the values: cold and hot reservoirs temperatures respectively are $\beta_C^{-1}=0.05$ and $\beta_H^{-1}=2$, $\lambda_1=0$ and $\lambda_2$ varies from $0$ to $4$. The blue line is for $\langle Q\rangle_{AB}$, yellow line for $\langle Q\rangle_{BC}$, green line for $\langle Q\rangle_{CD}$ and red line for $\langle Q\rangle_{DA}$.}
	\label{qty2}
\end{figure}

According to the efficiency definition (see Eq. (\ref{efic1})) it is calculated by $\eta=1-\frac{|\langle Q\rangle_{BC}|+|\langle Q\rangle_{CD}|}{\langle Q\rangle_{AB}+\langle Q\rangle_{DA}}$. 
 The heats exchanged at each part of the cycle are shown in Fig. \ref{qty2},
and it is evident from it that higher efficiency values are attained principally because the absolute value $|\langle Q\rangle_{BC}|$ is lower at the level crossing points. 

In Stirling cycle (see Fig. \ref{ciclo}), $|\langle Q\rangle_{BC}|$ represents one part of the heat released by the quantum heat engine, therefore, if it is lower then the work value increases, so increasing the efficiency. From Eq. (\ref{qbc}) we have $\langle Q\rangle_{BC}=U_C-U_B$, where its absolute value is lower if the value of $U_C$ is closer to $U_B$.
The difference between the energies $U_C$ and $U_B$ is caused by their different temperatures $\beta_C^{-1}$ and $\beta_H^{-1}$ respectively, 
then $U_C$ and $U_B$ will have close values each other when the principal energy level contributions to the mean energy $U$, i. e. the ground and first excited levels, are closer between them; such situation happens at $\lambda_2$ values around the degeneracy ground level points.
In summary, the modification of the energy level population at the process $BC$ of the cycle costs less heat when ground level crossings occur. 

Consequently, for low temperatures the ground state degeneracy becomes relevant to the increase of efficiency of our quantum heat engine. Such degeneracies are characteristic of quantum phase transition models, present respectively in the ferromagnetic phase of the Lipkin-Meshkov-Glick model, and in the superradiant phase of the full Dicke model.


\begin{thebibliography}{99}

\bibitem{Millen2016} James Millen and Andr\'e Xuereb, Perspective on quantum thermodynamics, {\it New. J. Phys.} {\bf 18} (2016) 011002.


\bibitem{Zeeya2017} Zeeya Merali, The new thermodynamics: how physics is bending the rules, {\it Nature} {\bf 551}
(2017) 20.

\bibitem{Alicki2018} Robert Alicki and Ronnie Kosloff, Introduction to Quantum Thermodynamics: History and Prospects {\it ArXiv} {\bf 1801.08314v2} (2018).

\bibitem{nielsen} Michael A. Nielsen and Isaac L. Chuang, {\it Quantum Computation and Quantum Information: 10th Anniversary Edition} (Cambridge University Press, 2011).


\bibitem{Alicki1979} Robert Alicki, The quantum open system as a model of the heat engine, {\it J. Phys. A: Math. Gen.} {\bf 12} (1979) L103.


\bibitem{Kieu2004}Tien D. Kieu, The Second Law, Maxwell's Demon, and Work Derivable from Quantum Heat Engines, {\it Phys. Rev. Lett.} {\bf 93} (2004) 140403-1.


\bibitem{niedenzu01}  Wolfgang Niedenzu {\it et al}., Quantum engine efficiency bound beyond the second law of thermodynamics, {\it Nature communications} {\bf 9} (2018) 165.

\bibitem{lutz1} J. Ro\ss{}nagel {\it et al}., Nanoscale Heat Engine Beyond the Carnot Limit {\it Phys. Rev. Lett.} {\bf 112} (2014) 030602.


\bibitem{niedenzu02} Wolfgang Niedenzu {\it et al}., On the operation of machines powered by quantum non-thermal baths, {\it New J. Phys.} {\bf 18} (2016) 083012.


\bibitem{squeezed01} Gonzalo Manzano {\it et al}., Entropy production and thermodynamic power of the squeezed thermal reservoir, {\it Phys. Rev. E} {\bf 93} (2016) 052120.

\bibitem{squeezed02} Jan Klaers {\it et al}., Squeezed Thermal Reservoirs as a Resource for a Nanomechanical Engine beyond the Carnot Limit, {\it Phys. Rev. X} {\bf 7} (2017) 031044.

\bibitem{squeezed03} Bo Xiao and Renfu Li, Finite time thermodynamic analysis of quantum Otto heat engine with squeezed thermal bath, {\it Phys. Lett. A} {\bf 382} (2018) 3051. 

\bibitem{su2018} Shanhe Su {\it et al}., The Carnot efficiency enabled by complete degeneracies, {\it Phys. Lett. A} {\bf 382} (2018) 2108.

\bibitem{Gardas}  Bart\l{}omiej Gardas and Sebastian Deffner,  Thermodynamic universality of quantum Carnot engines, {\it Phys. Rev. E} {\bf 92} (2015) 042126.

\bibitem{enrique2018}  Enrique Arias, Thiago R. de Oliveira and M. S. Sarandy , The Unruh quantum Otto engine, {\it J. High Energ. Phys.} {\bf 168} (2018).


\bibitem{expQHE01}  Johannes Ro{\ss}nagel {\it et al}.,  A single-atom heat engine, {\it Science} {\bf 352} (2016) 325.

\bibitem{expQHE02}  James Klatzow {\it et al}., Experimental Demonstration of Quantum Effects in the Operation of Microscopic Heat Engines, {\it Phys. Rev. Lett.} {\bf 122} (2019) 110601.

 \bibitem{collective00}  Hadrien Vroylandt, Massimiliano Esposito and Gatien Verley, Collective effects enhancing power and efficiency, {\it EPL (Europhysics Letters)} {\bf 120} (2017) 30009.

 \bibitem{collective01}  J. Jaramillo, M. Beau and A. del Campo, Quantum supremacy of many-particle thermal machines, {\it New J. Phys.} {\bf 18} (2016) 075019.


 \bibitem{collective02} Ali \"U. C. Hardal, Mauro  Paternostro and  \"Ozg\"ur E. M\"ustecapl\ifmmode \imath \else \i \fi{}o\ifmmode \breve{g}\else \u{g}\fi{}lu, Phase-space interference in extensive and nonextensive quantum heat engines, {\it Phys. Rev. E} {\bf 97} (2018) 042127.

 \bibitem{collective03} Thao P. Le {\it et al}., Spin-chain model of a many-body quantum battery, {\it Phys. Rev. A} {\bf 97} (2018) 022106.

 \bibitem{niedenzu03} Wolfgang Niedenzu and Gershon Kurizki, Cooperative many-body enhancement of quantum thermal machine power, {\it New J. Phys.} {\bf 20} (2018) 113038.

 \bibitem{sachdev} Subir Sachdev, {\it Quantum Phase Transitions} (Cambridge University Press, 2011).


 \bibitem{ma}  Yu-Han Ma, Shan-He Su and Chang-Pu Sun,  Quantum thermodynamic cycle with quantum phase transition, {\it Phys. Rev. E} {\bf 96} (2017) 022143.

  \bibitem{fazio}  Michele Campisi, and Rosario Fazio, The power of a critical heat engine, {\it Nature communications} {\bf 7} (2016) 11895. 

  \bibitem{kloc}  Michal Kloc, Pavel  Cejnar and Gernot Schaller, Collective performance of a finite-time quantum Otto cycle, {\it Phys. Rev. E} {\bf 100} (2019) 042126. 
  
  \bibitem{fadaie}  Mojde Fadaie, Elif Yunt and \"Ozg\"ur E. M\"ustecapl\ifmmode \imath \else \i \fi{}o\ifmmode \breve{g}\else \u{g}\fi{}lu, Topological phase transition in quantum-heat-engine cycles, {\it Phys. Rev. E} {\bf 98} (2018) 052124. 




  

   \bibitem{Dicke00} R. H. Dicke, Coherence in Spontaneous Radiation Processes, {\it Phys. Rev.} {\bf 93} (1954) 99. 

   \bibitem{Dicke01} Klaus Hepp and Elliott H. Lieb, On the superradiant phase transition for molecules in a quantized radiation field: the Dicke maser model, {\it Ann. Phys.} {\bf 76} (1973) 360.  

\bibitem{Dicke02} Clive  Emary and Tobias Brandes, Chaos and the quantum phase transition in the Dicke model, {\it Phys. Rev. E} {\bf 67} (2003) 066203.  

\bibitem{Aparicio2007} M. Aparicio Alcalde, A. L. L. de Lemos and N. F. Svaiter, Functional methods in the generalized Dicke model , {\it J. Phys. A: Math. Theor.} {\bf 40} (2007) 11961. 

\bibitem{Aparicio2009} M. Aparicio Alcalde, R. Kullock and N. F. Svaiter, Virtual Processes and Superradiance in Spin-Boson Models, {\it J. Math. Phys.} {\bf 50} (2009) 013511.

\bibitem{Aparicio2010} M. Aparicio Alcalde, A. H. Cardenas, N. F. Svaiter and V. B. Bezerra, Entangled states and superradiant phase transitions, {\it Phys. Rev. A} {\bf 81} (2010) 032335. 

\bibitem{Aparicio2011} M. Aparicio Alcalde, J. Stephany and N. F. Svaiter, Path integral approach to the full Dicke model with dipole-dipole interaction, {\it J. Phys. A: Math. Theor.} {\bf 44} (2011) 505301. 

   






\bibitem{LMGEntang01} Julien Vidal, Guillaume Palacios and R\'emy Mosseri, Entanglement in a second-order quantum phase transition, {\it Phys. Rev. A} {\bf 69} (2004) 022107.  


 \bibitem{DickeEntang01} N. Lambert, C. Emary and T. Brandes,  Entanglement and entropy in a spin-boson quantum phase transition, {\it Phys. Rev. A} {\bf 71} (2005) 053804.


 \bibitem{chaos-lmg1} D. C. Meredith, S. E. Koonin and M. R. Zirnbauer,  Quantum chaos in a schematic shell model, {\it Phys. Rev. A} {\bf 37} (1988) 3499.

  \bibitem{chaos-lmg2} Tobias Gra\ss{} {\it et al}., Quantum Chaos in SU(3) Models with Trapped Ions, {\it Phys. Rev. Lett.} {\bf 111} (2013) 090404.

  \bibitem{heiss01} W. D. Heiss, F. G. Scholtz and H. B. Geyer, The large $N$ behaviour of the Lipkin model and exceptional points, {\it  J. Phys. A: Math. Gen.} {\bf 38} (2005) 1843.

 \bibitem{LMG3} Pedro Ribeiro, Julien Vidal and R\'emy Mosseri, Exact spectrum of the Lipkin-Meshkov-Glick model in the thermodynamic limit and finite-size corrections, {\it Phys. Rev. E} {\bf 78} (2008) 021106.

  \bibitem{relano1} P. P\'erez-Fern\'andez {\it et al}., Quantum quench influenced by an excited-state phase transition, {\it Phys. Rev. A} {\bf 83} (2011) 033802.
  
  \bibitem{brandes} Tobias Brandes, Excited-state quantum phase transitions in Dicke superradiance models, {\it Phys. Rev. E} {\bf 88} (2013) 032133.
  
  \bibitem{baumann} Kristian  Baumann {\it et al}., Dicke quantum phase transition with a superfluid gas in an optical cavity, {\it Nature} {\bf 464} (2010) 1301.

  \bibitem{baden} Markus P. Baden {\it et al}., Realization of the Dicke Model Using Cavity-Assisted Raman Transitions, {\it Phys. Rev. Lett.} {\bf 113} (2014) 020408. 


\bibitem{fusco}  Lorenzo Fusco, Mauro Paternostro and Gabriele De Chiara, Work extraction and energy storage in the Dicke model, {\it Phys. Rev. E} {\bf 94} (2016) 052122.  


   \bibitem{QTDicke01}  Dario Ferraro {\it et al}., High-Power Collective Charging of a Solid-State Quantum Battery, {\it Phys. Rev. Lett} {\bf 120} (2018) 117702. 
   
\bibitem{Talkner2007} Peter Talkner, Eric Lutz and Peter H\"anggi, Fluctuation theorems: Work is not an observable, {\it Phys. Rev. E} {\bf 75} (2007) 050102.
 

\bibitem{Dicke05} M. Aparicio Alcalde {\it et al}., Thermal phase transitions for Dicke-type models in the ultrastrong-coupling limit, {\it Phys. Rev. E} {\bf 86} (2012) 012101. 

   \bibitem{Dicke03} F. t. Hioe, Phase Transitions in Some Generalized Dicke Models of Superradiance, {\it Phys. Rev. A} {\bf 8} (1973) 1440.
  
   \bibitem{Dicke04} M. Aparicio Alcalde and B.   M. Pimentel, Path integral approach to the full Dicke model, {\it Phys. A: Stat. Mech. Appl.} {\bf 390} (2011) 3385.

   
 
   \bibitem{cejnar} Pavel Cejnar and Pavel Stránský, Heat capacity for systems with excited-state quantum phase transitions, {\it Phys. Lett. A} {\bf 381} (2017) 984.  
   
   \bibitem{relano} P. P\'erez-Fern\'andez and A. Rela\~no, From thermal to excited-state quantum phase transition: The Dicke model, {\it Phys. Rev. E} {\bf 96} (2017) 012121.  

   \bibitem{popov} V. N. Popov and V. S. Yarunin, {\it Collective Effects in Quantum Statistics of Radiation and Matter} (Springer Netherlands, 1988).













\end{thebibliography}

\end{document}